# HePGA: A Heterogeneous Processing-in-Memory based GNN Training Accelerator

*Chukwufumnanya Ogbogu[1], Gaurav Narang[2], Biresh Kumar Joardar[3], Janardhan Rao Doppa[1], Krishnendu Chakrabarty[2], Partha Pratim Pande[1]
[1]Washington State University, Pullman, WA, USA. [2]Arizona State University, Tempe, AZ, USA. [3]University of Houston, Houston, TX, USA.

*Abstract*— Processing-In-Memory (PIM) architectures offer a promising approach to accelerate Graph Neural Network (GNN) training and inference. However, various PIM devices such as ReRAM, FeFET, PCM, MRAM, and SRAM exist, with each device offering unique trade-offs in terms of power, latency, area, and non-idealities. A heterogeneous manycore architecture enabled by 3D integration can combine multiple PIM devices on a single platform, to enable energy-efficient and high-performance GNN training. In this work, we propose a 3D heterogeneous PIM-based accelerator for GNN training referred to as HePGA. We leverage the unique characteristics of GNN layers and associated computing kernels to optimize their mapping on to different PIM devices as well as planar tiers. Our experimental analysis shows that HePGA outperforms existing PIM-based architectures by up to 3.8× and 6.8× in energy-efficiency (TOPS/W) and compute efficiency (TOPS/mm²) respectively, without sacrificing the GNN prediction accuracy. Finally, we demonstrate the applicability of HePGA to accelerate inferencing of emerging transformer models.

## I. INTRODUCTION

Graph neural networks (GNNs) have become the mainstream machine learning (ML) techniques for learning and decision-making from graph data, excelling in applications including recommendation systems, fraud detection, drug discovery, and traffic forecasting [1]. However, state-of-the-art GNN models are becoming increasingly complex, and real-world graph datasets are growing rapidly in size, consisting of millions of nodes and edges [2]. Hence, the compute- and data-intensive nature of GNN workloads leads to high-latency and energy costs when executed on traditional von-Neuman architectures (e.g., CPUs and GPUs) due to their limited memory bandwidth [3]. Consequently, there has been a significant demand for processing-in-memory (PIM)-based architectures that seamlessly integrate memory and computing to overcome this challenge by reducing data movement [3].

A GNN model is made up of several neural layers, with each layer comprising of an *aggregation* and a *combination* kernel [2]. During GNN training and inference, these kernels perform Matrix-Vector Multiplication (MVM) operations involving a graph adjacency matrix and a weight matrix in each layer, which makes PIM well-suited for executing GNN workloads [4] [5]. However, GNN layers have diverse computational and memory requirements, due to variations in the number of weight parameters, and the node features and the connectivity of graph datasets [5]. For example, the combination kernel in the first layer of a Graph Convolutional Network (GCN) model trained with the Amazon2M dataset has fewer parameters compared to the intermediate layers. Moreover, the MVM operations in the aggregation and combination kernels produce intermediate activations with varying computational and storage requirements across layers. These differences become profound in large-scale GNN workloads, where the size of the adjacency matrices differ in terms of sparsity, connectivity, and storage needs from one graph dataset to another. A single PIM device does not have all the ideal characteristics in terms of storage density, energy consumption, and robustness. Hence, this motivates the need for a heterogeneous manycore architecture involving multiple PIM devices to achieve the best trade-off in terms of power, area, performance, and GNN model accuracy while designing a suitable accelerator platform.

Selecting suitable PIM devices for a heterogeneous PIM-based architecture is challenging due to the wide range of memory technologies, including SRAM, DRAM, and non-volatile memory options including ReRAM, PCM, FeFET, and MRAM etc. [6] [7] [8]. Table I shows the characteristics of SRAM, FeFET and ReRAM memory technologies as examples. As shown in Table I, these devices have their distinct advantages and limitations in terms of energy, area, latency, and endurance. In addition, integrating different memory devices into a single monolithic die is not possible due to technological limitations [9]. However, three-dimensional (3D) integration can integrate diverse memory technologies in one platform [9].

Moreover, the complex interactions between the combination and aggregation phases of the GNN computational kernel generate multicast traffic that creates performance bottlenecks [10]. Hence, to alleviate these challenges, the mapping of the GNN layers and the associated compute kernels must minimize inter-layer communication overhead while meeting the thermal and power constraints. Overall, suitable mapping of the GNN layers to specific PIM-based Processing Elements (PEs) is crucial for achieving efficient data flow and in turn reduce latency and energy consumption during GNN training without compromising the prediction accuracy.

In this work, we propose a 3D heterogeneous PIM-based accelerator suited for GNN training referred to as **HePGA**. HePGA is enabled by a PIM device- and workload-aware design space exploration methodology for GNN training. Our proposed optimization framework jointly leverages the properties of the GNN layers and their kernels (such as bit-precision, number of weights, activations, and size of the adjacency matrix), as well as their recursive multicast data-flow patterns to determine an

TABLE I
COMPARISON OF VARIOUS PIM DEVICES

| Property | SRAM [8] | ReRAM [43] | FeFET [41] |
|---|---|---|---|
| Multi-bit Cell | No | Yes | Yes |
| †Cell Area ($F^2$) | 150 | 4 | 35 |
| Write Energy | 3pJ | 2nJ | 5pJ |
| Write Latency | ~1ns | ~100ns | ~3ns |
| Endurance(cycles) | >$10^{17}$ | $10^8$ | $10^5$ |
| Leakage Energy | High | Low | Low |

†F is the minimum feature size [8].

* Work done while at Washington State University
This work was supported, in part by the US National Science Foundation (NSF) under grants CNS-1955353, and CNS-1955196.

efficient mapping of GNN kernels to PIM-based PEs, and a suitable PE to planar tier mapping in the 3D heterogeneous architecture. In addition, we demonstrate the scalability of the HePGA architecture to diverse GNN models including Graph Convolutional Networks (GCN), Graph Sample Aggregate (SAGE) and Graph Attention Networks (GAT) [1] [11] [12]. We also show that the HePGA architecture can be extended beyond GNNs to support emerging transformer models. The key contributions of this work are:

- We propose a 3D heterogeneous manycore architecture referred to as HePGA, tailor-made for GNN training. HePGA leverages an optimized GNN layer-wise kernel mapping to achieve lower energy and better performance compared to existing homogeneous and heterogeneous architectures.

- We demonstrate the scalability of the HePGA architecture to a variety of graph datasets and GNN models for diverse tasks such as node classification and link prediction. We also demonstrate the applicability of the HePGA architecture to emerging transformer models.

- Our experimental results show that HePGA outperforms existing state-of-the-art architectures by up to 3.8× and 6.8× in terms of energy efficiency (TOPS/W) and area efficiency (TOPS/mm$^2$), respectively.

## II. RELATED PRIOR WORK

Graphics Processing Units (GPUs) are typically employed for accelerating GNN training and inference. However, GPUs suffer from low performance/watt and limited memory bandwidth. Domain-specific accelerators (e.g., EnGN, Rubik and I-GCN) alleviate these challenges [13] [14] [15]. However, they accelerate both the aggregation and combination kernels using the same compute engine, which is sub-optimal for handling the computing diversity in GNNs. On the contrary, HyGCN and FlowGNN leverage a specialized engine for accelerating each of the GNN compute kernels [16] [17]. However, a major drawback of these architectures is that they rely on separate memory units, requiring frequent data transfers between the accelerator and memory during GNN training. Prior work has proposed PIM-based GNN accelerators [10][18] [19]. ReMaGN is a homogeneous architecture that leverages ReRAMs for accelerating all the GNN layers [10]. As mentioned earlier, using one type of PIM device may lead to sub-optimal performance during GNN training. However, these architectures are still based on one type of PIM device and are not well-optimized for performing all the different types of compute kernels during GNN training. For example, the backpropagation operation during GNN training requires multiple crossbar write operations, which can cause the ReRAM cells to wear out quickly due to their limited write endurance. Other PIM-based architectures that leverage SRAMs and FeFET have been proposed for accelerating GNN workloads [5] [20]. However, these homogeneous architectures are built solely using either ReRAM, FeFET or SRAM PIM devices. HePGA outperforms these homogeneous architectures as shown later.

Prior work has proposed heterogeneous architectures that leverage more than one PIM device for accelerating ML workloads. HyperX is a heterogeneous architecture that leverages ReRAM- and SRAM-based crossbars to map static and trainable neural layers respectively for fine-tuning purposes

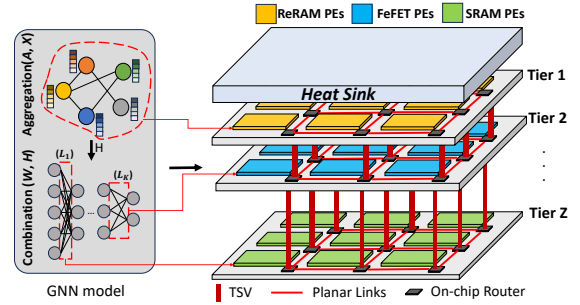

Fig. 1: Illustration of GNN layer-wise kernel to PE mapping and corresponding PE to 3D planar tier mapping. Here, the adjacency matrix ($A$) is mapped to ReRAM-based PEs and placed on Tier 1 as an example.

[21]. HyDe and HuNT propose design space exploration frameworks for finding an optimal mapping of convolutional neural network (CNN) layers to different PIM devices [22] [23]. They do not consider GNN training and more recent workloads such as Transformers. Hence, a suitable heterogeneous PIM-based architecture for GNN training and beyond needs to be explored. To fill this critical gap in the current state of art, we propose a heterogeneous PIM-based architecture (HePGA) that leverages the compute and communication characteristics of the GNN training to design a suitable hardware architecture.

## III. THE HEPGA ARCHITECTURE

### A. Problem Setup

Without loss of generality, we consider a GNN workload with $K$ neural layers, where each layer ($l$) consists of aggregation and combination kernels. As discussed earlier, these kernels rely on MVM operations involving an adjacency matrix ($A$), weight matrix ($W_l$) in the forward phase of GNN training, and activations ($H_l$). It is preferable to execute the MVM operations in each kernel on Non-Volatile memory (NVM)-based PIM. The number of weight parameters and activations vary across the GNN layers. Also, the activations are used in the high-precision calculation of gradients for weight updates in the backpropagation phase. These gradient calculations require multiple write operations, necessitating the need for a memory technology with high write endurance, such as SRAM [8]. Hence, a heterogeneous architecture with different PIM-based PEs can cater to the varying requirements of the GNN layers in both the forward and backpropagation phases.

We consider a manycore system with $C$ PIM-based PEs distributed across $Z$ planar tiers in a 3D architecture. Each tier contains PEs of a single type of PIM device. Fig. 1 illustrates an example of a GNN workload mapped on to a 3D heterogeneous architecture consisting of SRAM-, ReRAM-, and FeFET-based PEs. 3D architectures are prone to thermal hotspots, which causes variations in stored weights, adjacency matrix, and activations, especially in NVM devices (FeFET and ReRAM). This potentially leads to a degradation in the GNN prediction accuracy. Hence, the placement of the planar tiers (with a specific type of PIM-based PEs) with respect to heat sink impacts the accuracy. Given the varying (a) GNN layer characteristics, (b) differing requirements of forward and backpropagation phases, and (c) physical properties of PIM devices, our goal is to design a suitable heterogeneous manycore architecture (HePGA) and find an optimized layer-wise kernel mapping for a given GNN workload. To this end, we formulate a multi-objective optimization (MOO) problem to find the suitable mapping of GNN layer-wise kernels to PIM-based PEs, as well as their appropriate location in one of the $Z$ planar tiers, that achieves a suitable latency, area, power, and accuracy trade-off.

*B. HePGA Design & GNN Layers Mapping Methodology*

The inputs to the MOO framework include the GNN workload characteristics, i.e., graph adjacency matrix ($A$), $K$ neural layers with weights ($W_l$), activations ($H_l$), and architectural parameters such as the total number of PEs ($C$), PIM device options (e.g., $S|R|F$), and the number of planar tiers ($Z$). We define a GNN layer-wise kernel mapping ($\pi$) such that:

$$\pi = [A^{S|R|F}, W_1^{S|R|F}, H_1^{S|R|F}, \ldots, W_K^{S|R|F}, H_K^{S|R|F}] \quad (1)$$

where $A^{S|R|F}$, $W_l^{S|R|F}$, and $H_l^{S|R|F}$ represent the mapping of the adjacency matrix ($A$), weights ($W_l$) and activations ($H_l$) of the $l^{th}$ GNN layer respectively to either SRAM, ReRAM or FeFET-based PEs. Note that the adjacency matrix ($A$) of a given subgraph does not vary across the GNN layers ($l$). Also, we define a tier configuration ($\Gamma$) such that $\Gamma = [\Gamma_1^{S|R|F}, \ldots, \Gamma_Z^{S|R|F}]$, where $\Gamma_i^{S|R|F}$ represents the type of the PIM-based PEs in each planar tier in the 3D heterogeneous architecture. Here, the planar tiers $\Gamma_1^{S|R|F}$ and $\Gamma_Z^{S|R|F}$ are closest and farthest from the heat sink respectively, resulting in a higher temperature on the bottom tier compared to the tier closest to the heat sink. Our goal is to find an optimized mapping ($\pi, \Gamma$) such that we minimize (a) the loss in prediction **accuracy** ($Err$) due to temperature-induced PIM device non-idealities, (b) the **area** ($Ar$) in terms of the number of PEs needed to map all the GNN layers, (c) the **latency** ($Lat$), and (d) **power** consumption ($Pwr$). This MOO formulation is represented as:

$$D^* = MOO(Pwr(\pi,\Gamma), Ar(\pi,\Gamma), Lat(\pi,\Gamma), Err(\pi,\Gamma)) \quad (2)$$

where $D^*$ is the set of Pareto optimal designs ($\pi, \Gamma$ candidates) that achieve suitable trade-offs between the different design objectives. We find the optimized Pareto set of designs using the well-known MOO solver, AMOSA [24]. Next, we explain the evaluation of the design objectives: latency, area, accuracy, and power. Since we need to evaluate many design choices to solve the MOO problem, we consider surrogate design objectives elaborated below for tractable optimization instead of performing expensive cycle-accurate simulations.

**Latency ($Lat$):** Different PIM-based PEs (SRAM, ReRAM, FeFET) offer varying computational speeds while executing the GNN kernels. The multicast on-chip traffic pattern due to transfer of intermediate activations ($H_l$) and gradients ($\delta_l$) between PEs in the forward and backpropagation phases determine the communication latency. The overall latency ($Lat$) is influenced by both the computation ($\mathcal{L}_{comp}$) and communication ($\mathcal{L}_{comm}$) latencies, as shown in (3) and (4) below respectively. The communication latency, $\mathcal{L}_{comm}$ is determined by the Manhattan distance ($M_{ab}$) between two PEs ($PE_a$ and $PE_b$) as shown in (4), which is dictated by the mapping of $K$ GNN neural layers.

$$\mathcal{L}_{comp} \propto \sum_{l=1}^{K}[Latency(A^{S|R|F} + W_l^{S|R|F} + H_l^{S|R|F})] \quad (3)$$

$$\mathcal{L}_{comm} \propto \{H_l, \delta_l\} \cdot M_{ab} \quad (4)$$

**Area ($Ar$):** GNN training requires crossbars to store the layer-wise GNN weights ($W_l$), adjacency matrix ($A$) in forward phase, and activations ($H_l$) to compute gradients ($\delta_l$) in backpropagation phase. The total storage requirements for all the GNN layers depend on the type of PIM-based PEs to which they are mapped, as shown in (5).

$$Ar \propto \sum_{l=1}^{K}[Area(A^{S|R|F} + W_l^{S|R|F} + H_l^{S|R|F})] \quad (5)$$

**Accuracy ($Err$):** Variation in stored GNN parameters ($W_l$, $A$, and $H_l$) due to thermal noise varies with the type of PIM device. For instance, the conductance range (the gap between ON- and OFF-state conductance) in ReRAMs reduces exponentially with increase in temperature [25]. Meanwhile, in FeFET, the memory window (the separation between HIGH- and LOW-state threshold voltages) reduces linearly with temperature [7]. This reduction in conductance range/memory window results in loss in GNN prediction accuracy. In contrast, SRAMs are more robust to thermal noise compared to ReRAMs and FeFETs [22]. Further, thermal hotspots in a 3D architecture depend on the GNN layers' mapping and the power density of PEs on which these layers are mapped. For example, if multiple high power consuming PEs are stacked on top of each other, it can lead to higher thermal hotspots compared to when these PEs are placed far apart in a 3D architecture. Temperature of each PE ($Q_{o,z}$) depends on the PE's power consumption ($P_{o,u}$) as well as the architectural properties, given by (6) [26].

$$Q_{o,z} = \left\{\sum_{u=1}^{z}\left(P_{o,u}\sum_{v=1}^{u}R_v\right) + R_{bl}\sum_{u=1}^{z}P_{o,u}\right\} * \Phi_H \quad (6)$$

where $P_{o,u}$ is the power consumption of the PEs $u$ tiers away from the sink in a vertical stack $o$, $\Phi_H$ represents the lateral heat flow, $R_v$ is the thermal resistance in the vertical direction, and $R_{bl}$ is the thermal resistance of the base layer on which the die is placed, and $z$ represents the $z^{th}$ tier where PEs are located. Overall, the loss in accuracy ($Err$) due to thermal noise-induced non-idealities depends on the GNN kernels and their layer-to-PE mapping ($\pi$), and 3D tier configuration ($\Gamma$), as shown in (7), where $\mathcal{N}$ is the thermal noise of each PIM device type [25].

$$Err = \sum_{l=1}^{K}[\mathcal{N}(A^{S|R|F} + W_l^{S|R|F} + H_l^{S|R|F})] \quad (7)$$

**Power ($Pwr$):** Computation of GNN kernels on the PEs as well as multicast communication between the PEs lead to significant power consumption during GNN training. Here, computation power corresponds to the power incurred in the combination and aggregation phase of each GNN layer while mapped to either SRAM-, ReRAM- or FeFET-based PEs, as shown in (8). Further, data transfers such as intermediate activations ($H_l$) and gradients ($\delta_l$) between PEs in the forward and backpropagation phases result in significant power dissipation in the routers and links associated with the PEs. The communication power to transfer data ($H_l$ and $\delta_l$) between PEs is given by (9), where $M_{ab}$ is the corresponding Manhattan distance between the PEs, governed by the GNN layer mapping.

$$P_{compute} \propto \sum_{l=1}^{K}[Power(A^{S|R|F} + W_l^{S|R|F} + H_l^{S|R|F})] \quad (8)$$

$$P_{comm} \propto \{H_l, \delta_l\} \cdot M_{ab} \quad (9)$$

| Algorithm 1. HePGA design and GNN layer mapping optimization |
|---|
| **Input**: Manycore system with $C$ PEs of different PIM device type choices (SRAM, ReRAM or FeFET) |
| $APP$ = GNN training task |
| **Output**: $D^*$, the optimized Pareto set of designs (optimized GNN layer mapping and 3D planar tier configuration) |

| | |
|---|---|
| 1: | **Initialize**: $D$ = non-dominated set of solutions; $Av$ = Archive |
| 2: | Input variables $(\vec{x})$ = GNN layer-to-PE mapping $(\pi)$ and planar tier configuration $(\Gamma)$ |
| 3: | **Repeat**: |
| 4: | Select one $\vec{x}$ from $Av$ & *perturb* $\vec{x}$ to get a mapping $(\pi, \Gamma)$ |
| 5: | Evaluate$((\pi, \Gamma), APP)$ /∗ using (3)-(9) [section III B]∗/ |
| 6: | Update non-dominated set of solutions $D$ via $Pwr, Ar, Lat, Err$ |
| 7: | Update Archive $Av$ |
| 8: | **Until convergence or maximum iterations** |
| 9: | Pareto optimal set of designs $D^* \leftarrow D$ |
| 10: | **return** $D^*$, the optimized Pareto set of designs (optimized GNN layers mapping and HePGA architecture) |

***AMOSA-based MOO Approach***: We employ the well-known AMOSA solver to optimize GNN layer mapping on the HePGA architecture [24]. Algorithm 1 shows the high-level pseudocode for our design optimization methodology. The input variables $\vec{x}$ in our MOO approach are the GNN layer-wise kernel mapping $(\pi)$ and tier configuration $(\Gamma)$. First, we start with a randomly chosen mapping of layer-wise GNN kernels ($W_l$ and $A_l$) and activations ($H_l$) to PEs and 3D tier configuration. Next, we perturb a candidate mapping solution to get a new GNN layer mapping $(\pi)$ and tier configuration $(\Gamma)$ (Algorithm 1, line 4). In each optimization iteration, the selected mapping $(\pi, \Gamma)$ is evaluated for objectives using (3) to (9) (Algorithm 1, line 5). The non-dominated set of designs and $Archive$ are updated based on this new design and mapping evaluation (Algorithm 1, line 6 and 7). At convergence or after maximum iterations, we get the optimized Pareto set of designs $D^*$ from the MOO solver (Algorithm 1, line 10). Finally, we select the design from $D^*$ that achieves the best performance in terms of energy-efficiency (TOPS/W) and compute-efficiency (TOPS/mm$^2$) by performing cycle-accurate simulations. It should be noted that the GNN layer mapping to PEs and 3D tier configuration are determined at the design time (offline) for a given GNN.

## IV. EXPERIMENTAL RESULTS AND ANALYSIS

### A. Experimental Setup

The optimization of the HePGA architecture is executed on an NVIDIA V100 GPU with 32GB of memory. Algorithm 1 optimizes the GNN layer-to-PE $(\pi)$ and PE-to-tier mapping $(\Gamma)$ for HePGA architecture. It is executed for 100 iterations, as this is sufficient to ensure the convergence of the AMOSA-based MOO. The HePGA architecture considered in this work consists of 36 processing elements (PEs) as an example, distributed over four planar tiers and connected to each other using through-silicon-via (TSV) based vertical links. Our analysis and findings remain consistent regardless of the adopted system size. Table II presents the characteristics of the different PIM-based PEs considered and the physical parameters for the TSV. We consider an *iso-PE area* setting, where the different PIM-based PEs occupy the same area. However, the storage density of each PE differs from one device to another. The area, energy, and latency of the SRAM- FeFET- and ReRAM-based crossbars, as well as their peripheral circuits in each PE (shown in Table II), are modeled using NeuroSim [27]. Following prior work, we adopt a 3D Mesh as the interconnection topology, and standard NoC flow control mechanism (FIFO-based) for synchronization in the HePGA architecture [28]. The communication latency and the power dissipation due to inter-PE data exchange are estimated using BookSim [29]. Hotspot is used to estimate the on-chip temperature using the power traces from NeuroSim and BookSim [30]. Finally, we model the thermal effects on the GNN accuracy using the PyTorch wrapper in NeuroSim [27].

TABLE II
HEPGA SPECIFICATIONS

| 36 PEs distributed over 4 tiers (9 PEs/tier), 4 tiles/PE | |
|---|---|
| ReRAM Tile | 96 SAR ADCs (8-bits), 128×96 DACs (1-bit), 96 crossbars, 128×128 crossbar array, 2-bit/cell resolution, tech. node- 32nm, 0.33W, 0.40 $mm^2$ |
| FeFET Tile | 256×48 S/A (1-bit), 48 crossbars, 256×256 crossbar array, 1-bit/cell resolution, tech node- 28nm, 0.54W, 0.40$mm^2$ |
| SRAM Tile | 6T, 1-bit-cell, 256 S/A, 8KB SRAM array (256×256), 9 column/row-decoder, 9 SRAM arrays, tech node-14nm, 0.84W, 0.40 $mm^2$ |
| TSV Parameters | Diameter–5μm, Via Height–25μm, Capacitance – 37fF, Resistance– 20mΩ [42] |

The GNN models considered in this work include; GCN, GAT and SAGE [1] [12] [11]. Each GNN model considered uses four layers (i.e., $K = 4$), as an example. However, the proposed GNN layer mapping optimization technique is not limited by the layer count ($K$) and can be applied to GNNs with any number of layers. We evaluate the performance of HePGA using these GNN models on the PPI, Reddit (RDT), Amazon2M (A2M), and ogbl-citation2 (OC2) graph datasets [2]. These datasets are representative of real-world graph learning tasks such as node-classification and link prediction [2]. The large monolithic graphs are partitioned into multiple subgraphs using METIS to reduce memory requirements, following prior work [31] [32]. In addition, we leverage the well-known Reverse-Cuthill-Mckee (RCM) diagonalization technique to improve the graph adjacency matrix storage efficiency [33] [34]. We map multiple subgraphs simultaneously on HePGA depending on the size of the graph dataset to maximize throughput during the GNN training. We train each GNN model for 200 epochs. In our experimental evaluation, we consider energy-efficiency (TOPS/W) and area-efficiency (TOPS/mm$^2$) as the two relevant performance metrics that capture the latency, area, power objectives discussed above.

### B. HePGA Architecture & GNN Layer Mapping Trade-offs

The layer-to-PE mapping $(\pi)$ for a given GNN model, and the 3D tier configuration $(\Gamma)$ of HePGA influence the latency, area, power, and prediction accuracy. Fig. 2 (a) shows the Pareto front for the GCN model trained using the PPI dataset considering the abovementioned design objectives as an example. As shown in Fig. 2 (a), SRAM-based PEs are present in all the optimized Pareto configurations (highlighted in black) as they are preferred for gradient computation. Furthermore, SRAM-based PEs are mapped on the planar tier ($S_4$) farthest from the heat sink as they are more robust against thermal noise compared to the NVM devices, as shown in Fig. 3(b). It should be noted that to efficiently utilize on-chip hardware resources, we compute gradients layer-by-layer basis [27]. As shown in Fig. 2(a), the homogeneous architectures score high in one specific design objective while neglecting the others. For example, the configuration [$R_1R_2R_3R_4$], where all the planar tiers consist of only ReRAM-based devices, suffers significant loss in accuracy due to thermal non-idealities but has low area. The [$R_1R_2R_3R_4$] configuration is essentially the ReMaGN architecture proposed in prior work [10]. Fig. 2(b) shows the comparative performance evaluation of the optimized Pareto set of designs in terms of energy-efficiency (TOPS/W). As shown in Fig. 2(b), the tier configuration [$R_1R_2F_3S_4$] (highlighted in green) achieves the highest TOPS/W compared to the rest of the optimized Pareto designs. It utilizes ReRAM-based PEs (planar tier 1 and 2) and FeFET-based PEs (tier 3) for computations in the forward phase, and SRAM-based PEs (tier 4) for high-precision gradients ($\delta$) calculation in the backpropagation phase, as shown in Fig. 3(b). From here on, we refer to this tier configuration [$R_1R_2F_3S_4$] as **HePGA**.

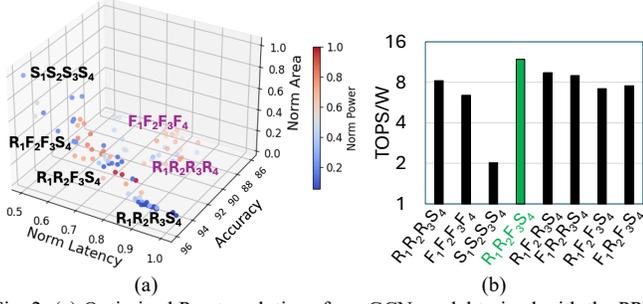

Fig. 2: (a) Optimized Pareto solutions for a GCN model trained with the PPI dataset. (b) Comparison of various Pareto optimal layer-to-PE mappings ($\pi$) and tier configurations ($\Gamma$) in terms of energy efficiency (TOPS/W).

The characteristics of the GNN layers and their associated kernels are crucial for determining layer mapping on the HePGA architecture to achieve high energy- and area-efficiency. The characteristics of the combination kernel vary across layers in terms of the number of weights and activations, while the aggregation kernel varies based on the graph dataset (adjacency matrix size for different subgraphs). Fig. 3(a) shows the layer-wise variation in terms of the number of crossbars required for the weights ($W_l$), activations ($H_l$) in the combination phase for a GCN model trained with the A2M dataset as a representative example. As shown in Fig. 3(a), it is worth noting that the characteristics of adjacency matrix ($A$) of a subgraph do not change across the layers. Fig. 3(b) illustrates the optimized mapping of the layer-wise kernels of the GCN model on the HePGA architecture ([$R_1R_2F_3S_4$]). Here, we observe that the GNN layers with high crossbar requirements for the weights and activations (e.g., $W_2$, $W_3$, $H_2$, and $H_3$), as well as the adjacency matrix ($A$) are mapped to ReRAM-based PEs, due to their higher storage density than FeFET. This reduces the inter-PE communication traffic as the layer-wise kernels are mapped to fewer PEs. Also, layers with fewer number of weights, but high activation counts (leading to high communication latency) are mapped on low-latency FeFET-based PEs. Furthermore, the weights and adjacency matrices are mapped to the neighboring PEs to minimize the communication latency and power, arising due to the inter-PE dataflow. This highlights the importance of considering the computation and communication characteristics of GNNs while finding an optimized layer mapping to achieve high energy- and area-efficiency.

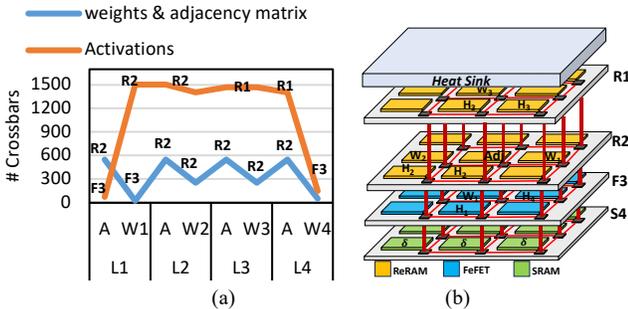

Fig. 3: (a) Variation in crossbar requirements, and mapping of the different kernels in layers ($L_1$-$L_4$) of the GCN model trained with the A2M dataset (b) Corresponding optimized mapping on the HePGA architecture ([$R_1R_2F_3S_4$]).

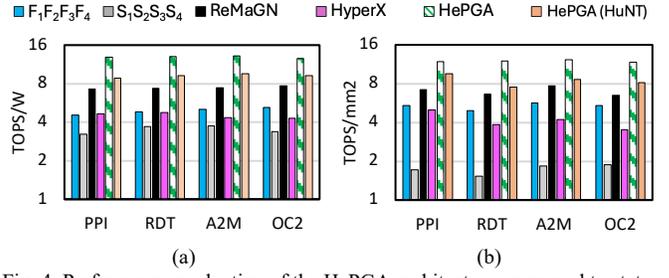

Fig. 4: Performance evaluation of the HePGA architecture compared to state-of-the-art homogeneous and heterogeneous counterparts in terms of (a) energy efficiency, and (b) area efficiency.

### C. Overall Performance Evaluation

Here, we compare HePGA's performance with existing architectures. Figs. 4(a) and 4(b) compare HePGA's energy-efficiency and area-efficiency with the state-of-the-art architectures during GCN training across various datasets. We compare HePGA with HyperX ([$R_1R_2S_3S_4$]), ReMaGN ([$R_1R_2R_3R_4$]), and homogeneous FeFET [$F_1F_2F_3F_4$]- and SRAM [$S_1S_2S_3S_4$]-based architectures. For brevity, we consider the GCN model as an example, however, a similar trend is observed for other types of GNN models. As shown in Figs. 4(a) and 4(b), HePGA achieves $2.6\times(2.2\times)$, $3.7\times(6.9\times)$, $1.7\times(1.7\times)$, $2.9\times(3\times)$ and $1.8\times(1.4\times)$ improvement in TOPS/W (TOPS/mm$^2$) on average over [$F_1F_2F_3F_4$], [$S_1S_2S_3S_4$], ReMaGN, and HyperX across the datasets respectively. HePGA exploits device heterogeneity and an optimized mapping of GNN layers to achieve the highest TOPS/W and TOPS/mm$^2$. Similar to GNNs, CNNs also have varying layer characteristics in terms of the number of parameters and their impact on accuracy and overall performance. As mentioned above, the recently proposed HuNT framework maps CNN workloads to a heterogeneous PIM-based architecture [23]. Hence, we also compare the performance of HePGA with a mapping used for only CNNs (HuNT) [23]. The HuNT-enabled mapping assumes a linear flow of activations between the adjacent neural layers and is oblivious to the multicast dataflow pattern inherent in GNN models. Hence, when GNN layers are mapped to HePGA using the HuNT mapping technique, we observe a degradation in performance due to the increased communication latency. Here, HePGA achieves up to $1.5\times(1.6\times)$ improvement in TOPS/W (TOPS/mm$^2$) compared to the HuNT-enabled mapping. As the HuNT-based mapping on HePGA is not appropriate for GNN workloads, we do not consider it for any further analysis.

Figs. 5(a) and 5(b) show the accuracy and endurance analysis of HePGA respectively. Here, endurance is measured in terms of the number of training instances that can be executed. HePGA achieves ~8% improvement on average in the GNN prediction accuracy compared to HyperX, as shown in Fig. 5(a). The SRAM-based homogeneous architecture achieves the highest accuracy due to its less vulnerability to thermal issues. HyperX has the lowest accuracy compared to other architectures, as the weights of some of the GNN layers are not updated while the others are trained [21]. As shown in Fig. 5(b), HePGA has a $100\times$ and $10^5\times$ higher number of possible training instances than ReMaGN and [$F_1F_2F_3F_4$] respectively, because it uses SRAM for backpropagation, while ReMaGN and [$F_1F_2F_3F_4$] rely solely on NVM-based PEs. Overall, HePGA offers the highest TOPS/W and TOPS/mm$^2$ with minimal GNN accuracy loss, while maintaining a good balance in endurance compared to other architectures.

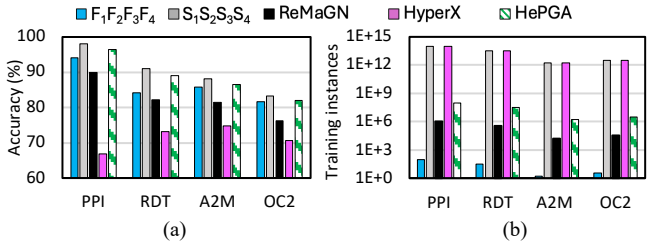

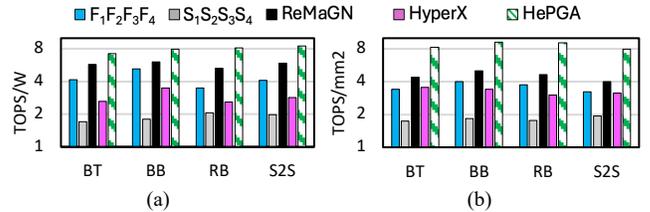

Fig. 5: Comparison of HePGA with existing architectures in terms of (a) GCN model accuracy, and (b) endurance (number of possible training instances).

Fig. 7: Comparison of HePGA with other architectures for accelerating Transformer inference in terms of (a) TOPS/W, and (b) TOPS/mm$^2$.

*D. Transferability of HePGA across GNN models*

In this subsection, we demonstrate that the HePGA architecture optimized using one type of GNN model is transferable to other GNNs. Specifically, we show that the optimized GNN layer mapping for the GCN model can also be used to map GAT and SAGE models to the HePGA architecture with minimal degradation in performance. Figs. 6(a) and 6(b) compare the GNN layer mapping for an optimized GNN model (SAGE and GAT), and their respective transferred models (GCN → SAGE and GCN→ GAT) in terms of energy-efficiency and accuracy. As shown in Figs. 6(a) and 6(b), we observe an average of less than 2%, and 1% drop in energy efficiency and accuracy respectively across the datasets. We observe a similar trend in the area-efficiency (TOPS/mm$^2$) across all GNN models and datasets. This demonstrates the transferability of the GNN layer mapping among various models, thereby reducing the cost of repeated optimizations for new GNN workloads.

*E. Applicability of HePGA to Transformers*

Transformer models have become the main-stream ML technique for natural language processing (NLP) and beyond [35]. In this work, we also demonstrate the applicability of HePGA to Transformer models. We have considered the BERT-Tiny (BT), BERT-Base (BB), RoBERTa (RB), and Seq2seq (S2S) models using SWAG QnA dataset as examples here [36] [37] [38]. In this work, we execute inference passes for these NLP tasks on the HePGA architecture. In contrast with CNN and GNN models, transformers introduce additional constraints on MVM operations with static and dynamic operands. Hence, the multi-head attention kernels of the transformer models that require high number of updates due to the dynamically changing input tokens are mapped to the SRAM tier ($S_4$). Other kernels that remain static during inference such as the feed-forward layers are mapped to the ReRAM- and FeFET-based tiers. Figs. 7(a) and 7(b) compare the performance of HePGA with existing counterparts in terms of TOPS/W and TOPS/mm$^2$ respectively, when they are used for accelerating the NLP models considered. Overall, we observe an improvement of up to $4.4\times$ and $5.2\times$ in TOPS/W and TOPS/mm$^2$ respectively over other architectures considered here. We also compare the performance of HePGA with state-of-the-art transformer accelerators; TransPIM and HAIMA [39] [40]. These architectures leverage DRAM and SRAM-based PEs for accelerating Transformers, which have high read/write latency and low storage density [40]. As a result, HePGA achieves an improvement of up to $5.5\times$ ($11.2\times$) and $1.5\times$ ($4.1\times$) in TOPS/W(TOPS/mm$^2$) over TransPIM and HAIMA respectively. This analysis demonstrates that the HePGA architecture is also suitable for transformer workloads.

## V. CONCLUSION

PIM-based architectures offer high-performance and energy-efficient hardware acceleration for GNN training. However, each PIM device has its own unique strengths and weaknesses. In this work, we propose a 3D heterogeneous architecture called HePGA that combines multiple PIM devices in a single system to meet the computation and communication demands of GNN training. Our approach leverages the varying characteristics of the GNN layers and associated kernels to map them on different PIM-based PEs and corresponding planar tiers. Overall, the HePGA architecture offers an improvement of up to $3.8\times$ ($6.8\times$) in TOPS/W(TOPS/mm$^2$) compared to homogeneous and existing heterogeneous PIM counterparts, without sacrificing prediction accuracy. Finally, we also demonstrate the applicability of HePGA to emerging transformer models. Our results show that HePGA is up to $5.5\times$ and $11.2\times$ more energy- and area-efficient respectively, than other architectures when executing transformers for NLP tasks.

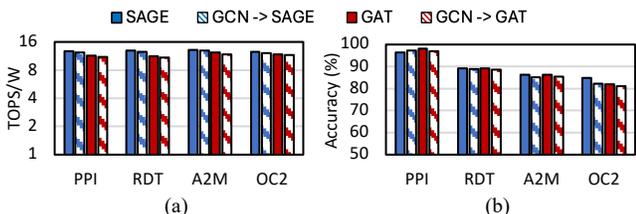

Fig. 6: Comparison of the optimized layer-wise kernel to PE mapping for SAGE and GAT models, and their transferred counterparts (GCN → SAGE and GCN→GAT) in terms of (a) TOPS/W, and (b) prediction accuracy.